# Extremophiles: a special or general case in the search for extra-terrestrial life?

**Ian von Hegner**
Aarhus University

**Abstract** Since time immemorial life has been viewed as fragile, yet over the past few decades it has been found that many extreme environments are inhabited by organisms known as extremophiles. Knowledge of their emergence, adaptability, and limitations seems to provide a guideline for the search of extra-terrestrial life, since some extremophiles presumably can survive in extreme environments such as Mars, Europa, and Enceladus. Due to physico-chemical constraints, the first life necessarily came into existence at the lower limit of it's conceivable complexity. Thus, the first life could not have been an extremophile, furthermore, since biological evolution occurs over time, then the dual knowledge regarding what specific extremophiles are capable of, and to the analogue environment on extreme worlds, will not be sufficient as a search criterion. This is because, even though an extremophile can live in an extreme environment here-and-now, its ancestor however could not live in that very same environment in the past, which means that no contemporary extremophiles exist in that environment. Furthermore, a theoretical framework should be able to predict whether extremophiles can be considered a special or general case in the galaxy. Thus, a question is raised: does Earth's continuous habitability represent an extreme or average value for planets? Thus, dependent on whether it is difficult or easy for worlds to maintain the habitability, the search for extra-terrestrial life with a focus on extremophiles will either represent a search for dying worlds, or a search for special life on living worlds, focusing too narrowly on extreme values.

**Introduction**
Life can reasonably well be considered fragile, and indeed, the fragility of most species is evident when they are taken away from their natural environment. In fact, one does not need to move very far beyond the protective confines of Earth's atmosphere, before life would cease to exist due to its encounter with the extreme cold of space, the vacuum and the intense solar radiation. However, over the past several decades, it has been found that many environments with extreme physico-chemical and climatic conditions are inhabited by diverse organisms known as extremophiles. These organisms, which were not known for a long time since such environments with aggressive parameters were assumed devoid of life, have continually expanded the knowledge of the limits of life [Pikuta et al., 2007].

Examples of such extremophiles include: *Deinococcus radiodurans*, a bacterium capable of surviving an acute dose of 5000 Gy of ionizing radiation and still recover as fully functional [Moseley and Mattingly, 1971]; *Pyrococcus furiosus*, an archaea which grows between 70 °C and 103 °C, with a preferred optimum temperature of 100 °C [Fiala and Stetter, 1986]; *Planococcus halocryophilus*, a bacterium which grows and divides at -15 °C and which is still metabolically active at -25 °C [Mykytczek et al., 2013], and *Chroococcidiopsis* sp., a cyanobacterium which can withstand several cycles of drying and wetting and capable of prolonged desiccation in extreme arid hot and cold deserts [Billi et al., 2001].

These and many other extremophiles have demonstrated that life *per se* is not fragile, and that the upper physical and chemical limits of terran life are in fact still not known [Harrison et al., 2013]. Indeed, in light of our knowledge of early Earth being an extreme environment, it has been conjectured that the first life was extremophilic, or, perhaps, hyperthermophilic [Pikuta et al., 2007].

Extremophiles have become an increasingly important field of research within astrobiology. One of the reasons for this is that these previously incognita organisms are considered relevant analogues of extra-terrestrial life that may exist within the Solar System [Martin and McMinn, 2018]. Indeed, while Earth is an oasis compared with other planets and moons in the Solar System, and much of terrestrial life cannot exist in those other places, it has increasingly been conjectured that some terrestrial extremophiles could probably survive in such worlds as Mars, Europa and Enceladus and thus, that extra-terrestrial analogues could exist there. Furthermore, the view that worlds with environments similar to those found on Earth, Mars, Europa, and Enceladus exist in large numbers in the galaxy is no longer controversial.

Thus, if extremophiles can be considered valid analogues of extra-terrestrial life, then our knowledge of extremophiles, their adaptability, and limitations could be a reasonable guide when searching for and theorizing the possibility of life elsewhere in the galaxy and beyond.

However, is the possibility of a planet, moon, exoplanet, or exomoons here-and-now to sustain extremophiles the same as the possibility that extremophiles could emerge and develop in these worlds? This can be questioned [von Hegner, 2019]. Let us consider the following statements: 'Φ is an extremophile', mean 'Φ descends from an extreme habitat.' Hence, 'Φ is an extremophile' but 'Φ descend not from an extreme habitat' is false as *per* definition; further, if 'Φ is a extremophile', 'Φ descends from an extreme habitat' and 'if Φ descends from an extreme habitat', 'Φ is a extremophile' are tautologies.

On the other hand, 'θ is inhabitable in the present' but 'θ was not inhabitable in the past' is not false *per* definition, because: (i) if 'θ is inhabitable in the present', 'θ was inhabitable in the past' and (ii) if 'θ was inhabitable in the past', 'θ is inhabitable in the present' are not mere tautologies.

Assuming life has not faced extinction by some natural catastrophe, then both statements would be valid if they were about Earth. However, when considering other worlds, then only statement (ii) would be valid. It might be possible for some terrestrial extremophiles to survive in e.g. Europa, if we transported them there, but it is not certain that they would have been able to emerge and evolve on that moon to begin with. Hence, contrary to what one can be tempted to believe, 'θ is habitable in the present' does not mean 'θ was habitable in the past'. Thus, being habitable in the present and being habitable in the past do not have the same semantic meaning in astrobiology.

Possessing a dual knowledge what specific terrestrial extremophiles are capable of and of the specific environment on a planet or moon does not necessarily mean that this dual knowledge can be used as a search criterion.

It is in fact two different situations, and in order to address this, it is necessary for a theoretical framework to incorporate not only prebiotic chemistry and planetary science, but also to incorporate evolutionary principles. A theoretical framework should also be able to give some predictions on whether extremophiles can be considered a special or general case in the galaxy. Both of these issues will be the subject of this article.

**Discussion**

While an extremophile can be defined as: "an organism that is capable of growth and reproduction within an environmental niche deemed detrimental to most life on Earth" [Martin and McMinn, 2018], or perhaps more adequately as "an organism that is tolerant to particular environmental extremes and that has evolved to grow optimally under one or more of these extreme conditions", implicit implying the existence of adaptive responses and survival thresholds to pressure, temperature, pH, salinity and desiccation extremes etc., a clarification of what life is will still be needed in order to understand these definitions.

There exists a variety of definitions of life. A modern and sufficient definition is as follows [von Hegner, 2019]: "Life$_{Terra}$ is a genome-containing, self-sustaining chemical dissipative system that maintains its localized level of organization at the expense of producing entropy in the environment; which has developed its numerous characteristics through pluripotential Darwinian evolution."

While it is easy to see that a hypothetical extremophile existing on, for example, the present-day Titan fits the first part of the definition, it is not easy however to see how this hypothetical extremophile fits the second part of the definition. Because, how should this extremophile have been able to undergo Darwinian evolution on this extreme moon of Saturn? What is important to realize here is that when we consider extra-terrestrial extremophiles, we are dealing with three closely entangled, yet still distinct situations that must be taken into consideration: chemical evolution, biological evolution, and the here-and-now.

*Chemical evolution*

If a planet possesses the right conditions for life, that is, a planet like the early Earth, then a process called OoL (Origins of Life) will presumably take place [Scharf et al., 2015]. This is the natural process or transition by which life will emerge from the synthesizing of the simplest organic building blocks from inorganic substrates, eventually leading to self-sustaining replicating systems.

Several lines of evidence point to the fact that life on Earth emerged 4.1–3.5 billion years ago [Bell et al., 2015], and there is some consensus regarding the assumption that the transition from chemistry to biology was not a single event, but rather a gradual series of events of increasing complexity [Scharf et al., 2015].

It is a debated question whether RNA or metabolism came first in this series [Domagal-Goldman et al., 2016]. However, regardless of whether the RNA-first model or the metabolism-first model is the correct one, it is clear that they both necessarily must have been integrated and trapped within a compartment before the emergence of the first fully autonomous cell [Domagal-Goldman et al., 2016].

*Biological evolution*

While the details are debated, and much remains to be elucidated, it is clear that prebiotic processes eventually lead to the emergence of the first fully autonomous cell, the descendents of which with a high certainty existed 3.5 billion years ago [Schopf et al., 2007]. Chemical evolution was presumably not a single event, but included multiple gradual processes that eventually united; biological evolution, however, began with a single event, namely when chemical evolution locked it on to this very first autonomous cell.

Gould (1996) put forward the Full House model in order to explain that although the phenomenon of increased complexity in the tree of life is evident, complex life arises only as a side consequence of a physico-chemical constrained starting point and actually represents only a relatively minor phenomenon. This statistical model focuses on the full system of variation as the reality, and explains that the portrayal of this complete system by a single figure construed as either the average or the extreme value within the system leads to an error. Instead, complexity represents only the small right tail of a bell curve of life with a constant mode at bacterial complexity. Thus, in biological evolution, there is no preferred direction in which organisms become more complex over time [Gould, 1996].

The Full House model posits that due to constraints imposed on the origins of life from chemical evolution and physical principles of self-organization, the first life form necessarily came into existence at the lower limit of life's conceivable and preservable complexity. Thus, the first life form is imposed to have begun with the simplest starting point right next to a lead wall of complexity [Gould, 1996]. This physico-chemical lower limit can be designated the 'left wall' for an architecture of minimal complexity. Since virtually no space exists between this left wall and the initial bacterial mode, nothing can move left. Only one spatio-temporal direction exists, which is toward increased complexity at the right [Gould, 1996].

While there is still much to be learned regarding chemical evolution, biological evolution is overall better understood. Bacteria represent the majority of life and it is well known that the root of the tree of life exists within their domain. Thus, the Archaea and Eukarya both share a common ancestry after their common lineage diverted from the bacterial mode [Ciccarelli et al., 2006].

Thus, biological evolution began with the first life, which is the simplest possible functional life. This may indeed be a universal starting point for life everywhere. This first life was according to this line of thought close to the bacterial mode and could thus have not been in the form of an extremophile or extremotolerant organism despite the fact that the early Earth overall evidently was an extreme environment compared with the present-day Earth.

Extremophiles are of course bacteria (or archaea) themselves. However, even though most bacteria evidently are resistant organisms in their environment, they are not *per se* extremophiles.

Thus, extremophiles do not specifically belong to the bacterial mode. They are more complex than the simplest bacteria and occupy the empty right region of complexity's space in environments, where the bacterial mode cannot exist. They extend the diversity range of life in the only accessible direction, that is, where the distribution of complexity becomes increasingly more right skewed through these occupations. The general life, that is, the bacterial mode, has increased in height and yet remained constant at their position in the full system of variation, from the emergence of life to the present day.

Extremophiles are *per* definition opposites to organisms, that are not extreme, and bacteria generally cannot exist in the very specific environments extremophiles have adapted to. Extremophiles thus descends from life, which must have been fragile in comparison with them.

*Extremophiles: the here-and-now*
As a result of physico-chemical necessities, life emerged as the simplest functional form, since there is no space for variation in the direction of less complexity [Gould, 1996]. Thermophiles are perhaps the oldest extremophiles on this planet. However, since the first life emerged at the lower limit of life's preservable complexity, they could not have been the first life, as previously discussed, since they are not the simplest conceivable organism. The first life would certainly have been resistant enough to exist in the specific microenvironment it emerged in, otherwise it would have disappeared again. However, this life, overall, would still have been too fragile to exist in the other environments outside its own microenvironment. The Hadean and first half of the Archean eras were overall extreme environments [Coenraads and Koivula, 2007].

Biological evolution is a stepwise process, wherein adaptations take place over time. Since extremophiles are more complex than the first forms of life, there must necessarily have been an intermediary duration of time between the first life and extremophiles where natural selection helped organisms adapt to their immediate environments.

However, although the first forms of life on Earth were not extremophiles, there is nevertheless strong evidence showing that extremophiles arose relatively early on Earth, and adaptation to high temperature might have been the first that life underwent. Model reconstruction of ancient RNA indicates that LUCA, the last universal common ancestor, was a thermophile or a hyperthermophile [Gaucher et al., 2010]. Furthermore, since the Earth mostly had extreme environmental conditions long into the Archean era, it still seems reasonable to use them to guide the search of life in other worlds.

However, as mentioned in the introduction: '$\theta$ is inhabitable" but "there is no life on $\theta$' is not false *per* definition, because: 'if $\theta$ is inhabitable', 'then there is life on $\theta$' and 'if there is life on $\theta$', '$\theta$ is inhabitable' are not mere tautologies. Hence, contrary to what one could believe, '$\theta$ is inhabitable' does not mean 'there is life on $\theta$'. Thus, being inhabitable and being inhabited do not have the same semantic meaning in astrobiology.

   This is due to the fact, that we are not considering only one situation, but two. It is true that some terrestrial extremophiles can presumably survive in another planet's or moon's extreme environment here-and-now. Yet, seemingly paradoxically, it is precisely this very extreme environment that ensures that in certain cases we can exclude the presence of extra-terrestrial extremophiles anyway.

   Extremophiles capable of living in nearly boiling water clearly exist in an extreme environment, where they evidently undergo biological evolution. However, this situation is for organisms that already exist in nearly boiling water. There exists an evolutionary continuum so to speak, in which adaptations take place, which implies a sequence, wherein organisms have adapted over time, and where the descendants have gained capabilities their ancestors did not possess. Thus, there has been a moment in the past where the ancestors of thermophiles, the first life, existed in water at a lower temperature, which means that a contemporary extremophile and its past ancestor did not possess the same capabilities. This means they could not have existed in the same environment. Thus, while there is a binary situation where extremophiles either exist here-and-now, or they do not, such a binary situation in terms of evolution is inadequate.

   An extremophile's ability to thrive in nearly boiling water is a capacity obtained through adaptation over time, wherein generations of organisms, step by step, have approached more and more extreme environments, where they adapted to the increasing temperatures. Thus, it is not only the physical parameters that must be taken into consideration, but the possibilities of biology must also be included. Thus, even though an extremophile can live in a planet's or moon's extreme environments here-and-now, its ancestor could not have lived in that very same environment in the past, which means, that no native contemporary extremophiles exist now in that environment.

When one considers, whether a current extremophile analogue can live on a given planet or moon, evolutionary principles must therefore be included. Apart from the simplest bacterial mode, there exist no 'platonic' species, that is, species that remain unchangeable over time. Mars, Europa, Enceladus, and perhaps Titan, all seem to possess the possibility to house extremophiles here-and-now. However, have these worlds ever during their geologic history possessed the possibility of adaptations over time? If the planet or moon always have been unfavorable for life overall, then life would not have been able to evolve over time, and extremophiles would not have emerged. Thus, a search for life on planets and moons that have always had an unfavorable environment for life, will be pointless.

   It could of course be pointed out that if lithopanspermia has merit, then extremophiles from a planet like Earth could be transported to another world, where they could not have originated, but where they thanks to their time in an evolutionary continuum on Earth can now exist. This is a valid but different point, which does not influence the specific situation discussed here.

   It could also be pointed out that if life as discussed emerged in safe microenvironments on an otherwise extreme Earth, and extremophiles arose relatively quickly on it, the same could be the case in other worlds. This is a valid point as well, however that is not the argument developed here. The argument being made is that the difference between worlds with and without a relaxed microenvironment is a profound one.

   Thus, the first life on a planet or moon must have existed in a more relaxed environment; a microenvironment. Even if it is conjectured that chemical evolution can lead to life under stable extreme conditions, that is, not a microenvironment, then its product–a fully autonomous cell which would experience several extreme physico-chemical factors simultaneously, would not be able to undergo biological evolution and become resistant or tolerant, if it had not been protected first against the multiple aggressive environmental factors. It would be destroyed the moment chemical evolution achieved it. So this is fundamentally not a question of the chemistry of life's origin; it is a question of the biology of life.

*Extremophiles: a special or general case?*
Terrestrial extremophiles evidently exist in abundance on a planet with a rich diversity of life. They are part of a vast interconnected ecosystem. However, extremophiles live on the edge of possibilities for life. Thus, they are in that sense always an expression of special life on Earth. If there indeed exist extremophile analogues now, for example, on Mars, then they represent life that has been cut off from a large ecosystem, wherein general life once existed, but no longer does.

The Noachian is an era on Mars that took place 4.1 to 3.7 billion years ago. This is approximately equivalent to the Hadean and early Archean eras on Earth, when the first terrestrial life forms emerged [Bell, et al., 2015]. During the Noachian, the atmosphere was likely denser than at present, and the planet may either have been relatively warm and wet, or cold with melting ice [Fastook and Head, 2015]. Some amount of past liquid water on the planet's surface is indicated by several lines of evidence, such as the existence of ancient, water-eroded structures, and weathered exhumed phyllosilicates [Carter et al., 2010]. It may even be the case that an ocean of liquid water covered the northern plains [Brandenburg, 1987]. There has presumably also existed a magnetic field [Connerney et al., 1999] that would have functioned as a protective shield against solar radiation and galactic cosmic rays.

If extra-terrestrial extremophiles exist on Mars now, then the planet must once have possessed a viable ecosystem that disappeared when global climate changes resulted in a colder, dryer, and hostile environment and the planet transitioned into the Hesperian and Amazonian eras. Thus, a search specifically for extremophile analogues on planets like Mars represents a search after extra-terrestrial life in worlds that from a biological point of view, are dying worlds.

It is correct that there do exist extremophiles apparently living isolated deep inside Earth's interior [Labonte et al., 2015]. However, *Candidatus Desulforudis audaxviator* evidently lives still in the Earth, a planet with an extensive viable ecosystem from which they themselves originated. Furthermore, a true isolated system only exists as a theoretical abstraction.

It is not yet known, what the probability is for the emergence of life in worlds with the right conditions, or what the correct conditions exactly are. So far, we only have one example of life: terrestrial life. That a planet has the right conditions for life is not the same as that life will emerge there. Many events in biology are due to contingent events, which are independent of biological evolution itself, and which historically could have proceeded differently. Whether there is a direction in chemical evolution, that is, that with time and the right conditions life will unavoidably emerge, is still an unanswered question.

An equally important question essential for the discussion here is, what the probability regarding an inhabited world maintaining that life is. Thus, it is not known what the probability is regarding worlds where life has emerged in the right conditions, can maintain these very right conditions.[1] We now have information regarding thousands of exoplanets and hundreds of Solar Systems [Schneider, 2019] but only information of one Solar System in which life has emerged so far. Therefore, presuming that life emerge elsewhere, it is not known what the distribution is for worlds where life has emerged. Life has existed on Earth for 3.5 billion years. However, is this an extreme value, or is it within the average value for planets on which life has emerged?

---

[1] I will here downplay the fact, that a star's luminosity changes over time, generally increasing [Rushby et al., 2013], and that a star gradually will move away from its position in the H-R-diagram [Green et al., 2004], which means, that the habitable zone moves outward, and that life on a planet, and the existence of the planet, is therefore limited in time. I will only consider the period as on Earth, where life emerged and evolved, and until this life disappears again when the star of the planet changes dramatically following its normal cycle. This is taken here as the general condition for all planets.

Earth has undergone several episodes, during which life became severely stressed. One example is the global Huronian glaciation event that occurred between 2.45 and 2.22 billion years ago and which was probably triggered by the rise of atmospheric oxygen and decreased methane, the so-called Great Oxygenation Event [Bekker, 2014]. Another example is the hypothesized Snowball Earth wherein the planet's surface became nearly completely covered with ice [Hoffman et al., 1998], with glacial ice sheets reaching the equator at least twice between 717 and 635 million years ago, with the glacial period in northwest Canada lasting ~55 million years [Rooney et al., 2014]. This latter condition became almost permanent. Earth barely seemed to become free from it again. However, life persisted through this, and the planet itself recovered. However, was this an average scenario, or is the Earth's survival an extremity?

If life emerged independently on both Earth and Mars and Earth maintained its ecosystem, whereas the ecosystem on Mars collapsed, then one could argue that the probability for planets to maintain life throughout their stars lifetimes is $p = 50\%$. However, this would be too simple, restricted only to two planets in a single Solar System.

Therefore, the question is whether the maintenance of habitability on Earth belongs to an extreme value, or whether Mars with its hypothetical life, even if planets like it exist in the habitable zone, belongs to the average scenario. More data from more planets and Solar Systems are required in order to make a frequency distribution that displays the true frequency of outcomes in a sample of Solar Systems where life has emerged. Nevertheless, understood this way, this hypothesis provides the following two possibilities:

(i) It means that if planets where life emerged, in general, had difficulty maintaining this habitability, as may be the case for Mars, then the search for extra-terrestrial life guided by a focus on extremophiles overall will be a search after dying worlds with an overall destroyed ecosystem from a biological point of view.

The utilization of extremophiles to guide the search for extra-terrestrial life is thus representative of the main class, as planets where life emerges, overall cannot maintain the conditions for life for long in geologic terms. Thus, extremophiles will be the last surviving class of life, fighting to maintain their existence in local habitats on planets that slowly lose the ability to sustain even these classes.

(ii) It means that if planets where life emerged, in general, easily maintain this habitability, as Earth seems to have done, then the search for extra-terrestrial life guided by a focus on extremophiles will be a special, and not a general search after life on living worlds, where one would focus too narrowly on extreme values instead of the full spectrum of variation in the entire ecosystem.

The application of extremophiles to guide the search for extra-terrestrial life is thus representative of only a subclass, because these organisms are only one part of the complexity distribution. Focusing only on these would overlook the entire picture, because general life on inhabited planets is the main class.

A simple equation to express this situation can be obtained using the following modified Drake equation:

$$I = R_* \cdot f_p \cdot f_l \cdot f_m \cdot T$$

where:

$I$ = the number of inhabited worlds in the galaxy,
$R_*$ = the average rate of star formation in the galaxy,
$f_p$ = the fraction of those stars that have planets,
$f_l$ = the fraction of planets (or moons) where life emerged and evolved,
$f_m$ = the fraction of inhabited planets (or moons) that make it through natural catastrophes throughout the lifetime of their star,

assuming, that $f_m$ is modest:

$T$ = the length of time for which extremophiles exist on dying inhabited planets or moons.

Understood this way, it is almost analogous to the study of the tail on a salamander, which has been cast off, but continues to move. The tail is undeniably biology, but it is dying. The general life is the salamander, which continues to live, and whose existence represents a viable ecosystem. This analogy is, of course, only partially correct. The priority in the search of extra-terrestrial life is without doubt the finding of extra-terrestrial life, regardless of whether it is general life, special life, or fossils. However, assuming life is common in the universe and we start locating life at different locations in the galaxy, then it will also be relevant whether it is general or special life that is located, how the distribution of inhabited worlds is, and how wide the search should be.

Earth and Mars can thus each represent the norm for inhabited worlds. However, it might be stated that there exists a type of inhabited world different from both Earth and Mars, which puts this into doubt, namely a world like Europa or Enceladus.

Europa is one of the four Galilean moons of Jupiter. It has an outer crust of solid ice and presumably a deep global ocean of liquid water beneath its surface [Kivelson et al., 2000] likely about 100 km deep [Chyba and Phillips, 2002], which, thanks to the heat from tidal dissipation, has remained in liquid form over geologic time [Cassen et al., 1980]. Likewise, Enceladus is a moon of Saturn that likewise presumably has a global subsurface liquid ocean beneath the frozen surface [Thomas et al. 2016] where also tidal dissipation driven by its orbit is the probable mechanism in the formation and maintenance of a liquid water layer over geologic times [Travis and Schubert 2015].

Worlds like these might represent plausible environments, where extra-terrestrial life once may have emerged in a relaxed favorable microenvironment, but where the surrounding extreme environment, in contrast to Earth's overall environment, did not relax afterward over time, but have remained extreme. However, at the same time they can represent worlds where the extreme environment has not exceeded a tipping point, as on Mars, and where the ecosystems are not dying. In such a situation, all life has apparently evolved into extremophiles, and the original bacterial mode they originated from may have disappeared again, as the microenvironment it emerged from may have disappeared as well. However, such a world, while not an Earth-like planet, does actually fall within the same class as Earth with regard to life.

First, while the environment in e.g. Europa might be considered extreme, it is still probably not entirely homogeneous. It would be sufficiently inhomogeneous for biological variation to take place there, and thus selection pressures for different extremophile analogues would exist. As previously discussed, life is constrained to begin at a point, in which the complexity distribution is bounded to one side. Life begins as the simplest living organism, and there is no space to vary in the direction of decreasing complexity at that starting point [Gould, 1996]. This means that extremophile analogues, which are not the simplest possible organisms, do not necessarily move in the complexity direction, but also occasionally toward simplicity in an unbiased random walk.

Thus, extremophile analogues can also lead to an ecosystem with a rich diversity of extremophile organisms with greater or less complexity within the parameters of the extreme environment.

Second, we can actually imagine two different worlds, perhaps exemplified by Europa and Enceladus. As in the above discussion, in one world we would encounter a world with a top extremophile average as on Earth, and another world where we would not. Even though such different worlds may have a continuously extreme environments and continuous extremophile analogues existing there, it can be predicted that we eventually would not encounter life that distinguishes itself as extremophile analogues in such an environment anyway. On Earth, extremophiles are distinctive by standing out from general life: they have a top extremophile average (that is, the single extremophile that initially stood out from the total distribution of extremophiles) in comparison with life as a whole. However, this does not have to be the case in worlds like Europa or Enceladus. Here, life with more uniformity would also be possible, yet the decline of the top extremophile average in such a world will not imply that there has been a decline in extremophile analogues adapting to the extreme environments.

This is due to the phenomenon that the variance of the top extremophile average decreases as the extremophile analogues overall get better and bette adapted to their environments. This would cause the extreme value of the distribution, that is, the top extremophile average, to decrease as well, even though this original top extremophile paradoxically would still exist. Moreover, since extremophiles by definition are in opposition to life, that is not extremophiles, it would not be meaningful to designate this ecosystem consisting only of extremophile analogues for extremophiles. Instead, they should be designated as general life.

However, one may wonder why the top extremophile analogues have not continued their increased adaptation? The answer is that they simply cannot do it (this is not necessarily the case for precisely these moons, but can be considered a general example).

Just as there is a left wall of minimal complexity for unicellular life, there is also a right wall of possibility for life [Gould, 1996]. This means that there is a limit on how far life can adapt to an extreme environment, even when there is sufficient time and opportunities for adaptation. The extremophile analogues that first reached the top extremophile average are now virtually standing still at the right wall of possibility, while the rest of the extremophile analogues gradually adapt further and further, finally reaching up to them. In other words, the top extremophile average is the right tail of a shrinking bell curve of extremophile averages with a stable mean. This is a consequence of overall improvements in capabilities.

The upper physico-chemical limit for potential extremophiles abilities is still unknown [Harrison et al., 2013]. The last decades have revealed many surprises, and extremophiles will presumably continue offering many surprises in the future. However, it is still highly probable that no extremophile ever will be able to adapt to withstanding a temperature corresponding to the inside of a type G star or the pressure corresponding to a neutron star. These predictions are not merely due to a limited and terrestrial biased focus on life as we know it. Instead, it is fundamentally a matter of physics, which imposes limits on what biomatter can achieve, representing a right wall of possibility.

So even worlds with hypothetical life like Europa or Enceladus will fall under the examined scenarios discussed for Earth or Mars. Have they undergone an upheaval in their ecosystems with consequences as on Mars, life will in the form of extremophile analogues either could exist in different microenvironments or disappear entirely. Have they made it through with an ecosystem, as on Earth, then they will be able to develop extremophile analogues or simply life.

**Conclusion**
The points discussed in this article are not intended as a criticism of the relevance of extremophiles as biological reference organisms in astrobiology. However, the search for extra-terrestrial life might be impeded since a framework linking life's capacity for evolutionary adaptation with environmental conditions past and present is lacking. Astrobiology is the offspring of both the Copernican and the Darwinian revolutions. Thus, what is needed is the characterization of the theoretical and experimental biological limits for the emergence, distribution, and limits for extremophiles and providing a framework guided by insights from planetary science and evolutionary biology.

If the goal is to find analogues to terrestrial extremophiles on extreme planets and moons, then we know that since the first life on Earth was not an extremophile, and that biological evolution necessitates a duration of time, the dual knowledge of what specific terrestrial extremophiles are capable of and the analogue environment on extreme planets or moons are only the first steps. The real consideration here is not whether terrestrial extremophiles can exist in the analogue environment, but instead whether the planet or moon has continuously had the same overall extreme environment since its formation, that is, whether the ancestral population of the extremophile analogues could emerge and evolve there in the past. Only if it had a favorable environment, or microenvironment, will extremophiles be a relevant search criterion.

Furthermore, when we formulate a framework for a planet's capacity for emergence of life, not only must the probability for life emerging on worlds with the right conditions be taken into consideration, but also the probabilities for such inhabited worlds maintaining this life throughout the life time of the worlds' stars. It is not yet known what this potential for planets in maintaining the right conditions for life is, but depending on whether planets like Earth represent an extreme or average value, or whether planets like Mars are representative, planets like Snowball Earth and present day Mars can be more frequent than planets like present day Earth, if it goes that life emerges on all, but that they afterwards generally cannot maintain a viable ecosystem. What is remarkable about extremophiles, if they do indeed exist on Mars is that they represent a form of life that holds on to a world which is dying from an ecosystem point of view. This means that extremophiles should be considered general cases in the search for extra-terrestrial life. If planets on the other hand have it easy in maintaining this inhabitability, then extremophiles will be special cases.

The characterization of these theoretical biological limits discussed have thus in some ways restricted the application of extremophiles as guidelines in the search for extra-terrestrial life, while in other regards have opened up for interesting scenarios regarding how the nature of the distribution of inhabited worlds can be in the galaxy. For now, we do not have enough data to answer these questions, since so far only one single Solar System with life is known. However, defining the theoretical restrictions and the experimental limits of inhabitability in terrestrial and extra-terrestrial environments will provide a valuable guideline for this particular area of astrobiology.